\documentclass[prl,showpacs,preprintnumbers,superscriptaddress,floatfix,onecolumn,amsmath,amssymb]{revtex4}
\usepackage{epsfig}
\usepackage{graphicx}
\usepackage{dcolumn}
\usepackage{bm}
\usepackage[active]{srcltx}

\newcommand{\be}{\begin{equation}}

\newcommand{\ee}{\end{equation}}
\newcommand{\bea}{\begin{eqnarray}}
\newcommand{\eea}{\end{eqnarray}}

\begin{document}

\preprint{}
\title{Synergy and redundancy in the Granger causal analysis of dynamical networks}
\date{\today}
\author{Sebastiano Stramaglia}
\affiliation{Center of Innovative Technologies for Signal Detection
and Processing TIRES, Dipartimento di Fisica, Universit\'a di Bari,
Italy\\} \affiliation{Computational Neuroimaging Lab. Biocruces
Health Research Institute. \\ Cruces University Hospital. Barakaldo,
Spain} \affiliation{Ikerbasque, The Basque Foundation for Science.
Bilbao, Spain}

\author{Jesus M Cortes}
\affiliation{Computational Neuroimaging Lab. Biocruces Health Research
Institute. \\ Cruces University Hospital. Barakaldo, Spain}
\affiliation{Ikerbasque, The Basque Foundation for Science. Bilbao, Spain}

\author{Daniele Marinazzo}
\affiliation{Faculty of Psychology and Educational Sciences,
Department of Data Analysis, Ghent University, Henri Dunantlaan 1,
B-9000. Ghent, Belgium}

\date{\today}

\begin{abstract}
We analyze by means of Granger causality the effect of synergy and redundancy in the
inference (from time series data) of the information flow between
subsystems of a complex network. Whilst we show that fully conditioned Granger causality is not
affected by synergy,  the
pairwise analysis fails to put in evidence synergetic effects.
 In cases when the number of samples is low,
thus making the fully conditioned approach unfeasible, we show that
partially conditioned Granger causality is an effective approach if
the set of  conditioning variables is properly chosen. We consider
here two different strategies (based either on informational
content for the candidate driver or on selecting the variables with
highest pairwise influences) for partially conditioned Granger
causality and show that depending on the data structure either one
or the other might be  valid. On the other hand, we observe
that fully conditioned approaches do not work well in presence of
redundancy, thus suggesting the strategy of separating the pairwise
links in two subsets: those corresponding to indirect connections of
the fully conditioned Granger causality (which should thus be excluded) and links that can be ascribed to redundancy effects and, together with the results from the fully
connected approach, provide a better description of the causality
pattern in presence of redundancy. We finally apply these methods to
two different real datasets. First, analyzing electrophysiological data from an epileptic brain, we show that synergetic effects are
dominant just before seizure
occurrences. Second, our analysis applied to gene expression time series
from  HeLa culture shows  that the underlying regulatory
networks are characterized by both redundancy and synergy.

\end{abstract}

\maketitle

\section{Introduction}
Living organisms can be modeled as an ensemble of complex
physiological systems, each with its own regulatory mechanism and
all continuously interacting between them  \cite{ivanov}. Therefore
inferring the interactions within and between these
modules is a crucial issue. Over the last years the interaction
structure of many complex systems has been mapped in terms of
networks, which have been successfully studied using tools from
statistical physics \cite{barabasi}. Dynamical networks  have
modeled physiological behavior in many applications; examples range
from networks of neurons \cite{netnn}, genetic networks
\cite{gennet}, protein interaction nets \cite{protein} and metabolic
networks \cite{metabolic,metabolic1,cortes2011}.

The inference of dynamical networks from time series data is related
to the estimation of the  information flow between variables
\cite{pereda2005}; see also \cite{book1,book2}. Granger causality
(GC) \cite{granger,sethbressler} has emerged as a major tool to
address this issue. This approach is based on prediction: if the
prediction error of the first time series is reduced by including
measurements from the second one in the linear regression model,
then the second time series is said to have a Granger causal
influence on the first one. It has been shown that GC is equivalent
to transfer entropy \cite{schreiber} in the Gaussian approximation
\cite{barnett} and for other distributions \cite{hac}. See
\cite{vicente} for a discussion about applicability of this notion
in neuroscience, and \cite{njp} for a discussion on the reliability of GC for continuous dynamical processes.
It is worth stressing that several forms of coupling may mediate information flow in the brain, see  \cite{bart1,bart2}.
The combination of GC and complex networks theory is also a promising line of research \cite{njp1}.

The pairwise Granger analysis (PWGC) consists in assessing
GC between each pair of variables, independently of the rest
of the system. It is well known that the pairwise analysis cannot
disambiguate direct and indirect interactions among
variables. The most straightforward extension, the
conditioning approach \cite{geweke}, removes indirect influences by evaluating to which extent the predictive
power of the driver on the target decreases when the conditioning variable is removed. It has to be noted however that, even though its limitations are well known, the pairwise GC approach is still used in situations
where the number of samples is limited and a fully conditioned
approach is unfeasible. As a convenient alternative to this suboptimal solution, a partially conditioned approach,
consisting in conditioning on a small number of variables,
chosen as the most informative ones for the driver node, has been proposed \cite{partialold}; this
approach leads to results very close to those obtained
with a fully conditioned analysis and even more accurate in the presence
of a small number of samples \cite{partialBC}. We remark that the
use of partially conditioned Granger causality (PCGC) may be useful
also in non-stationary conditions, where the GC pattern has to
be estimated on short time windows.

Sometimes though a fully conditioned (CGC) approach can encounter
conceptual limitations, on top of the practical and computational
ones: in the presence of redundant variables the application of the
standard analysis leads to underestimation of influences
\cite{noipre}. Redundancy and synergy are intuitive yet elusive
concepts, which have been investigated in different fields, from pure
information theory \cite{griffith,harder,Lizier}, to machine learning
\cite{yang} and neural systems \cite{schnei,gat},
with definitions that range from the purely operative to the most
conceptual ones. When analyzing interactions in multivariate time
series, redundancy may arise if some channels are all influenced by
another signal that is not included in the regression; another
source of redundancy may be the emergence of synchronization in a
subgroup of variables, without the need of an external influence.
Redundancy manifests itself through a high degree of correlation in
a group of variables, both for instantaneous and lagged influences.
Several approaches have been proposed in order to reduce
dimensionality in multivariate sets eliminating redundancy, relying
on generalized variance \cite{barnettpre}, principal components
analysis \cite{zhou}, or Granger causality itself \cite{noipla}.

A complementary concept to redundancy is synergy. The synergetic
effects that we address here, related to the analysis of dynamical influences
in multivariate time series, are similar to those encountered in
sociological and psychological modeling, where {\it suppressors} is
the name given to variables that increase the predictive validity
of another variable after  its inclusion into a linear regression
equation \cite{conger}; see \cite{thompson} for examples of easily
explainable suppressor variables in multiple regression research. Redundancy and synergy have been further connected to information transfer in \cite{preExpansion},where an expansion of the information flow has been
proposed to put in evidence redundant and synergetic multiplets of
variables. Other information-based approaches have also addressed
the issue of collective influence \cite{Chicharro,Lizier}. The
purpose of this paper is to provide evidence that in addition to the
problem related to indirect influence, PWGC shows another relevant
pitfall: it fails to detect synergetic effects in the information
flow, in other words it does not account for the presence of subsets
of variables that provide some information about the future of a
given target only when all the variables are used in the regression
model. We remark that since it processes the system as a whole, CGC
evidences synergetic effects; when the number of samples is low,
PCGC can detect synergetic effects too, after an adequate
selection of the conditioning variables.

The paper is organized as follows. In the next section we briefly
recall the concepts of GC and PCGC. In section III we describe some
toy systems illustrating how redundancy can affect the results of
CGC, whilst indirect interactions and synergy are the main problems
inherent to PWGC. In section IV we provide evidence of synergetic
effects in epilepsy: we analyze electroencephalographic recordings
from an epileptic patient corresponding to ten seconds before the
seizure onset; we show that the two contacts which constitute the putative seizure onset act as synergetic variables driving the rest of
the system. The pattern of information transfer evidences the
actual seizure onset only when synergy is correctly considered.

In section V we propose an approach that combines PWGC and CGC to
evidence the pairwise influences due only to redundancy and not
recognized by CGC. The conditioned GC pattern is used to partition
the pairwise links in two sets: those which are indirect influences
between the measured variables, according to CGC, and those which
are not explained as indirect relationships. The unexplained
pairwise links, presumably due to redundancy, are thus retained to
complement the information transfer pattern discovered by CGC. In
cases where the number of samples is so low that a fully
multivariate approach is unfeasible, PCGC may be applied instead of
CGC. We also address here the issue of  variables selection for
PCGC, and consider a novel strategy for the selection of variables:
for each target variable, one selects the variables sending the highest
amount of information to that node as indicated by a pairwise
analysis. By construction, this new selection strategy works more
efficiently when the interaction graph has a tree structure: indeed in this case
conditioning on the parents of the target node ensures that indirect
influences will be removed. In the epilepsy example  the selection
based on the mutual information with the candidate driver provides
results closer to those obtained by CGC. We finally apply the proposed approach on time series
of gene expressions, extracted from a data-set from the HeLa
culture. Section VI summarizes our conclusions.

\section{Insights into Granger causality}
Granger causality is a powerful and widespread data-driven  approach
to determine whether and how two time series exert direct dynamical
influences on each other \cite{hla}. A convenient nonlinear
generalization of GC has been implemented in \cite{noipre2},
exploiting the kernel trick, which makes computation of dot products in
high-dimensional feature spaces possible using simple functions (kernels)
defined on pairs of input patterns. This trick allows the
formulation of nonlinear variants of any algorithm that can be cast
in terms of dot products, for example Support Vector Machines
\cite{vapnik}. Hence in \cite{noipre2} the idea is still to perform
linear Granger causality, but in a space defined by the nonlinear
features of the data. This projection is conveniently and implicitly
performed through kernel functions \cite{shawe} and a statistical
procedure is used to avoid over-fitting.

Quantitatively, let us consider $n$ time series $\{x_\alpha
(t)\}_{\alpha =1,\ldots,n}$; the lagged state vectors are denoted
$$X_\alpha (t)= \left(x_\alpha (t-m),\ldots,x_\alpha (t-1)\right),$$
$m$ being the order of the model (window length). Let $\epsilon
\left(x_\alpha |{\bf X}\right)$ be the mean squared error prediction
of $x_\alpha$ on the basis of all the vectors ${\bf
X}=\{X_\beta\}_{\beta =1}^n$ (corresponding to the kernel approach
described in \cite{noiprl}). The multivariate Granger causality
index $\delta_{mv} (\beta \to \alpha )$ is defined as follows:
consider the prediction of $x_\alpha$ on the basis of all the
variables but $X_\beta$ and the prediction of $x_\alpha$ using all
the variables, then the GC is the (normalized) variation of the
error in the two conditions, i.e.
\begin{equation}\label{mv}
\delta_{mv} (\beta \to \alpha )= \log{\epsilon \left(x_\alpha |{\bf
X}\setminus X_\beta\right)\over \epsilon \left(x_\alpha |{\bf
X}\right)};
\end{equation}
In \cite{ancona} it has been shown that not all the kernels are
suitable to estimate GC. Two important classes of kernels
which can be used to construct nonlinear GC measures are the
{\it inhomogeneous polynomial kernel} (whose features are all the
monomials in the input variables up to the $p$-th degree; $p=1$
corresponds to linear Granger causality) and the {\it Gaussian
kernel}.

The pairwise Granger causality is given by:
\begin{equation}\label{bv}
\delta_{bv} (\beta \to \alpha )= \log{\epsilon \left(x_\alpha
|X_\alpha \right)\over \epsilon \left(x_\alpha |X_\alpha ,
X_\beta\right)}.
\end{equation}
The partially conditioned Granger causality is defined as follows. Let
$\bf{Y}$ be the variables in $\bf{X}$, excluding $X_\alpha$ and
$X_\beta$, then (\ref{mv}) can be written as:
\begin{equation}\label{pc}
\delta_{c}^{\bf{Y}} (\beta \to \alpha )= \log{\epsilon
\left(x_\alpha |X_\alpha , \bf{Y}\right)\over \epsilon
\left(x_\alpha |X_\alpha, X_\beta ,\bf{Y}\right)}.
\end{equation}
When  $\bf{Y}$ is only a subset of the total number of variables in
$\bf{X}$ not containing $X_\alpha$ and $X_\beta$, then
$\delta_{c}^{\bf{Y}}$ is called the partially conditioned Granger causality (PCGC). In \cite{partialold} the set $\bf{Y}$ is chosen as the
most informative for $X_\beta$. Here we will also consider an
alternative strategy: fixing a small number $k$, we select ${\bf Y}=
\{X_\gamma\}_{\gamma =1}^k$ as the $k$ variables with the maximal
pairwise GC $\delta_{bv} (\gamma \to \alpha )$ w.r.t. that target
node, excluding $X_\beta$.

\section{Examples}
In this section we provide some typical examples to remark possible
problems that pairwise and fully conditioned analysis may
encounter.
\subsection{Indirect GC among measured variables}
\begin{figure}[ht]
\includegraphics[width=12cm]{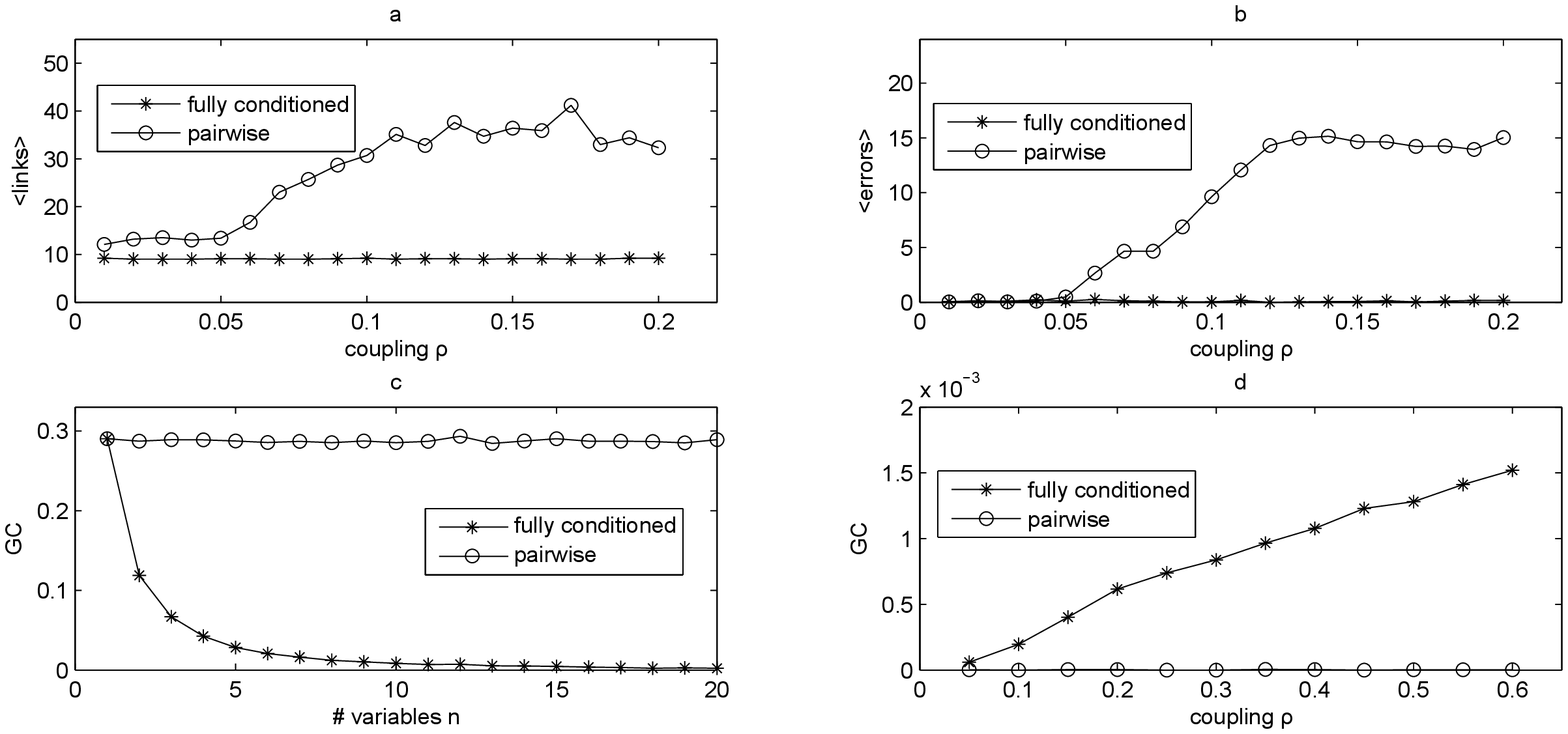}
\caption{{\rm  (a) The average number of links as a function of the
coupling $\rho$, over 100 runs of 2000 time points, retrieved by
PWGC and CGC on the coupled map lattice described in the text, eqs.
(\ref{catena1}-\ref{catena2}). (b) On the coupled map lattice, the
average error (sum of type I errors and type II errors in the
recovery of causal interactions) by PCGC, obtained by the IB
strategy and by PB, is plotted versus the coupling $\rho$. Errors
are averaged over 100 runs of 2000 time points. (c) For the example
dealing with redundancy, CGC and PWGC are plotted versus the number
of variables. Results are averaged over 100 runs of 1000 time points
(d) For the example dealing with synergy, CGC and PWGC are plotted
versus the coupling $\rho$. Results are averaged over 100 runs of
1000 time points
 \label{fig1}}}\end{figure}

We consider the following lattice of ten unidirectionally coupled noisy
logistic maps, with
\begin{equation}\label{catena1}
x_1 (t)=f\left( x_1(t-1)\right)+ 0.01 \eta_1 (t),
\end{equation}
and
\begin{equation}\label{catena2}
x_i (t)=(1-\rho)f\left( x_{i}(t-1)\right)+\rho f\left(
x_{i-1}(t-1)\right) + 0.01 \eta_i (t),
\end{equation}
with $i=2,\ldots,10$. Variables $\eta$ are unit variance Gaussian
noise terms. The transfer function is given by $f(x)=1-1.8 x^2$. In
this system the first map is evolving autonomously, whilst the other
maps are influenced by the previous ones with coupling $\rho$, thus
forming a cascade of interactions. In figure
 \ref{fig1}a  we plot as a function of $\rho$ the number of GC interactions found by
PWGC and CGC, using the method described in \cite{noipre} with the
inhomogeneous polynomial kernel of degree two. The CGC output is the
correct one (nine links) whilst the PWGC output also accounts for
indirect influences and therefore fails to provide the underlying
network of interactions. On this example we have also tested PCGC,
see figure \ref{fig1}b. We considered just one conditioning
variable, chosen according to the two strategies described above.
Firstly we consider the most informative w.r.t. the candidate
driver, as described in \cite{partialold}; we call this strategy
{\it information based} (IB). Secondly, we choose the variable characterized by the
maximal pairwise influence to the target node, a {\it pairwise
based} (PB) rule. The PB strategy yields the correct result in this
example, whilst the IB one fails when only one conditioning variable
is used and requires more than one conditioning variables to provide
the correct output. This occurrence is due to the tree topology of
the interactions in this example, which favors PB selecting by
construction the parents of each node.

\subsection{Redundancy due to a hidden source}
We show here how   redundancy constitutes a problem for CGC. Let
$h(t)$ be a zero mean and unit variance hidden Gaussian variable,
influencing $n$ variables $x_i (t)=h(t-1)+ s \eta_i (t)$, and let
$w(t)=h(t-2) +s \eta_0 (t)$ be another variable who is influenced by
$h$ but with a larger delay. The $\{\eta\}$ variables are unit
variance Gaussian noise and s controls the noise level. In figure
\ref{fig1}c we depict both the linear PWGC and the linear CGC from
one of the x's to w (note that h is not used in the regression
model). As $n$ increases, the conditioned GC vanishes as a
consequence of redundancy. The GC relation which is found in the
pairwise analysis is not revealed by CGC  because $\{x\}$ variables
are maximally correlated and thus $x_i$ drives $w$ only in the
absence of any other variables.

The correct way to describe the information flow pattern in this
example, where the true underlying source $h$ is unknown, is that
all the $\{x\}$ variables are sending the same information to w,
i.e. that variables $\{x\}$ constitute a redundant multiplet w.r.t.
the causal influence to w. This pattern follows from observing that for all x's CGC vanishes whilst PWGC does not
vanish. This example shows that, in presence of redundancy, the CGC
pattern alone is not sufficient to describe the information flow
pattern of the system, and also PWGC should be taken into account.
\subsection{Synergetic contributions}
Let us consider three unit variance  iid Gaussian noise terms $x_1$,
$x_2$ and $x_3$. Let $$x_4(t)=0.1 (x_1(t-1) + x_2(t-1))+\rho
x_2(t-1)x_3(t-1) + 0.1 \eta(t).$$ Considering the influence $3\to
4$, the CGC reveals that $3$ is influencing $4$, whilst PWGC fails
to detect this causal relationship, see figure \ref{fig1}d, where we
use the method described in \cite{noipre} with the inhomogeneous
polynomial kernel of degree two; $x_2$ is a suppressor variable for $x_3$ w.r.t. the influence on $x_4$.
This example shows that PWGC fails
to detect synergetic contributions. We remark that use of nonlinear GC is mandatory in this case to put in evidence the synergy between $x_2$ and $x_3$.
\subsection{Redundancy due to synchronization}
\begin{figure}[ht]
\includegraphics[width=13cm]{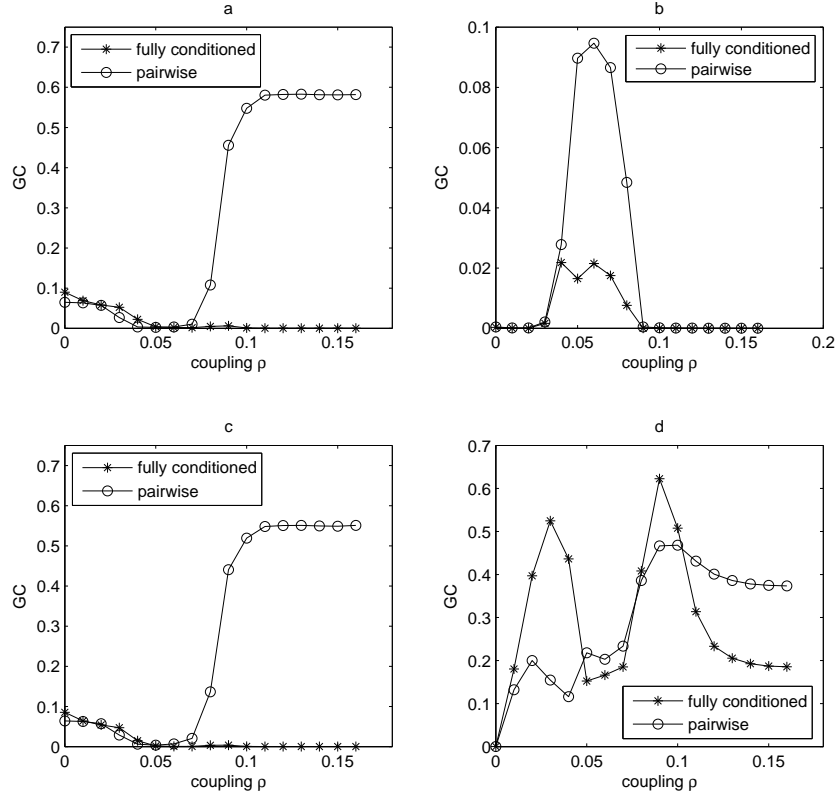}
\caption{{\rm  (a) The CGC and PWGC of the causal interaction $x_1
\to x_5$ is plotted as a function of the coupling $\rho$ for the
example dealing with redundancy due to synchronization, eqs.
(\ref{synch1}-\ref{synch2}). The linear algorithms are used here and
results are averaged over 100 runs of 2000 time points (b) The CGC
and PWGC of the causal interaction between two variables in the
multiplet is plotted as a function of the coupling $\rho$ for the
example dealing with redundancy due to synchronization. The linear
algorithms are used here and results are averaged over 100 runs of
2000 time points (c) The nonlinear CGC and PWGC of the causal
interaction $x_1 \to x_5$ is plotted as a function of the coupling
$\rho$ for the example dealing with redundancy due to
synchronization. The algorithm with the polynomial kernel of order 2
is used here and results are averaged over 100 runs of 2000 time
points (d) The nonlinear CGC and PWGC of the causal interaction
between two variables in the multiplet is plotted as a function of
the coupling $\rho$ for the example dealing with redundancy due to
synchronization. The algorithm with the polynomial kernel of order 2
is used here and results are averaged over 100 runs of 2000 time
points. \label{fig2}}}\end{figure}

As another example, we consider a toy system made of five variables
$\{x_i\}$. The first four constitute a multiplet made of  a fully
coupled lattice of noisy logistic maps with coupling $\rho$,
evolving independently of the fifth. The fifth variable is
influenced by the mean field from the coupled map lattice. The
equations are, for $i=1,\ldots,4$,:
\begin{equation}\label{synch1}
x_i (t)=(1-\rho)f\left( x_{i}(t-1)\right)+\rho \sum_{j=1,j\ne i}^4
f\left( x_{j}(t-1)\right) + 0.01 \eta_i (t),
\end{equation}
 and
 \begin{equation}\label{synch2} x_5
(t)=\sum_{i=1}^4 {x_i(t-1)\over 8}+ \eta_5 (t),
\end{equation}
where $\eta$ are unit variance Gaussian noise terms. Increasing the
coupling $\rho$ among the variables in the multiplet
$\{x_1,x_2,x_3,x_4\}$, the degree of synchronization among these variables (measured e.g. by  Pearson correlations)
increases and they become almost synchronized for  $\rho$ greater
than 0.1 (complete synchronization cannot be achieved due to the
noise terms); redundancy, in this example, arises due to complex
inherent dynamics of the units. In figure \ref{fig2} we depict both
the causality from one variable in the multiplet ($x_1$; the same results hold fo r$x_2$, $x_3$ and $x_4$ )  to
$x_5$, and the causality between pairs of variables in the
multiplet: both linear and nonlinear PWGC and CGC are shown for the
two quantities.

Concerning the causality $x_1 \to x_5$, we note that, for low
coupling, both PWGC and CGC, with linear or nonlinear kernel,
correctly detect the causal interaction. Around the transition to
synchronization, in a window centered at $\rho=0.05$, all the
algorithms fail to detect the causality $x_1 \to x_5$. In the {\it
almost} synchronized regime, $\rho > 0.1$, the fully conditioned
approach continues to fail due to redundancy, whilst the PWGC
provides correctly the causal influence, both using the linear and
the nonlinear algorithm.

As far as the causal interactions within the multiplet are
concerned, we note that using the linear approach we get small
values of causality just at the transition, whilst we get zero
values far from the transition. Using the nonlinear algorithm, which
is the correct one in this example as the system is nonlinear, we
obtain nonzero causality among the variables in the multiplet, using
both PWGC or CGC: the resulting curves are non-monotonous  as one
may expect due to the inherent nonlinear dynamics. For $\rho > 1$ nonzero GC is observed because of the noise which prevents the system to go in the complete synchronized state.

This example again shows that in presence of redundancy one should
take into account both CGC and PWGC results. Moreover it also shows
how nonlinearity may render extremely difficult the inference of
interactions: in this system there is a range of values,
corresponding to the onset of synchronization, in which all methods
fail to provide the correct causal interaction.

\section{Synergetic effects in the epileptic brain}
\begin{figure}[ht]
\includegraphics[width=13cm]{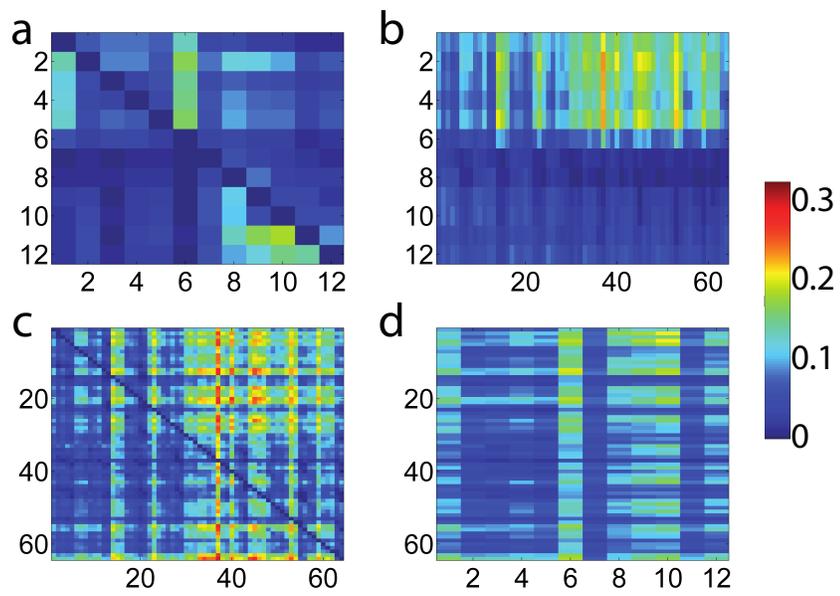}
\caption{{\rm The PWGC is depicted for the epilepsy data. (a) PWGC
between the contacts of the two DEs. (b) PWGC from DEs to CEs. (c)
PWGC between CEs. (d) PWGC from CEs to DEs.
 \label{fig3}}}\end{figure}

\begin{figure}[ht]
\includegraphics[width=13cm]{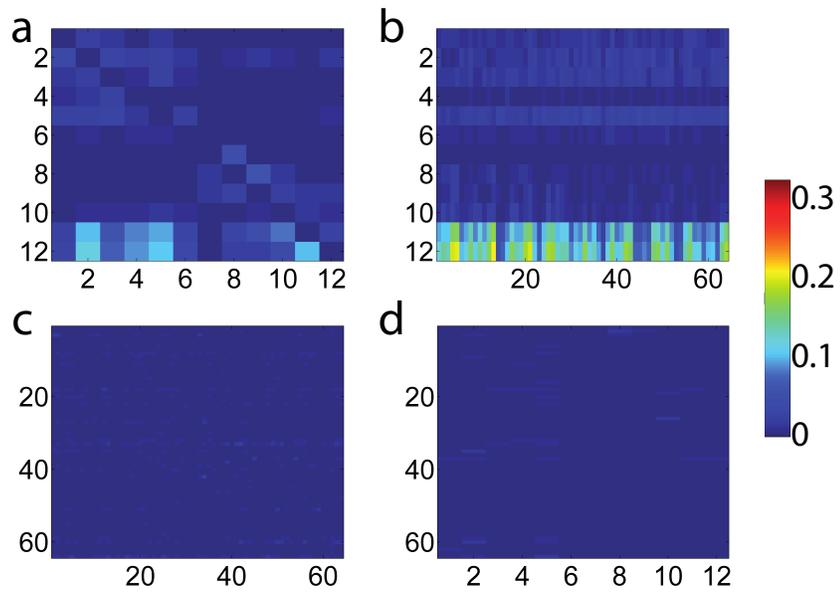}
\caption{{\rm The CGC is depicted for the epilepsy data. (a) CGC
between the contacts of the two DEs. (b) CGC from DEs to CEs. (c)
CGC between  CEs. (d) CGC from CEs to DEs.
 \label{fig4}}}\end{figure}

As a real example we consider intracranial EEG recordings from a
patient with drug-resistant epilepsy with an implanted array of
$8\times 8$ cortical electrodes (CE) and two depth electrodes (DE)
with six contacts each. The data are available at  \cite{kol_web}
and further details on the dataset are given in  \cite{kol_paper}.
Data were sampled at 400 Hz. We consider here a portion of data recorded
in the preictal period, 10 seconds preceding the seizure
onset. To handle this data, we use linear Granger causality with $m$
equal to five. In figure \ref{fig3} we depict the PWGC between DEs
(panel a), from DEs to CEs (panel b), between CEs (panel c) and from
CEs to DEs (panel d). We note a complex pattern of bivariate
interactions among CEs, whilst the first DE seems to be the
subcortical drive to the cortex. We remark that there is no PWGC
from the last two contacts of the second DE (channels 11 and 12) to
CEs and neither to the contacts of the first DE. In figure
\ref{fig4} we depict the CGC among DEs (panel a), from DEs to CEs
(panel b), among CEs (panel c) and from CEs to DEs (panel d). The
scenario in the conditioned case is clear: the contacts 11 and 12,
from the second DE, are the drivers both for the cortex and for the
subcortical region associated to the first DE. These two contacts
can be then associated to the seizure onset zone (SOZ). The high
pairwise GC strength among CEs is due to redundancy, as these latter are all
driven by the same subcortical source. Since the contact 12 is also
driving the contact 11, see figure \ref{fig4}a, we conclude that the
contact 12 is the closest to the SOZ, and that the contact 11 is a
{\it suppressor} variable for it, because it is necessary to include
it in the regression model  to put in evidence the influence of 12
on the rest of the system. Conversely, the contact 12 acts as a
suppressor for contact 11. We stress that the influence from
contacts 11 and 12 to the rest of the system emerges only in the CGC
and it is neglected by PWGC: these variables are synergetically
influencing the dynamics of the system. To our knowledge this is the
first time that synergetic effects are found in relation with
epilepsy.

\begin{figure}[ht]
\includegraphics[width=13cm]{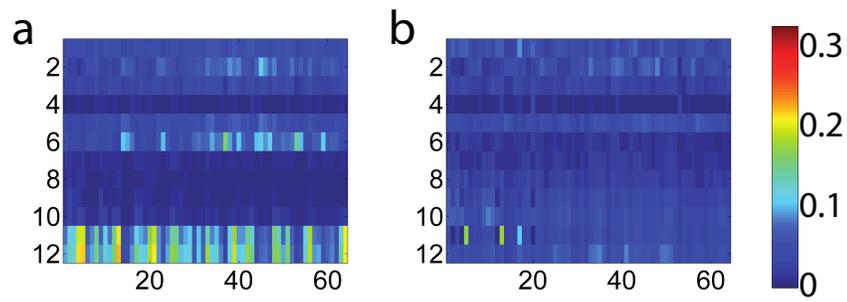}
\caption{{\rm The PCGC is depicted for the epilepsy data. (a)PCGC
from DEs to CEs, with IB strategy for variable selection. Note that the influences from DE11 are conditioned here on DE12, and the influences from DE11 are conditioned on DE 12, thus showing that these variables are suppressor among themselves.   (b) PCGC
from DEs to CEs, with PB strategy for variable selection.
 \label{fig5}}}\end{figure}

On this data we also apply PCGC  using one conditioning variable.
The results are depicted in figure \ref{fig5}: using the IB strategy
we obtain a pattern very close to the one from CGC, while this is
not the case of PB. These results seem to suggest that IB works
better in presence of redundancy, however we have not arguments to
claim that this a general rule. It is worth mentioning that in presence
of synergy the selection of variables for partial conditioning is
equivalent to the search of suppressor variables.

\section{A combination of pairwise and conditioned Granger Causality}
In the last sections we have shown that CGC encounters issues resulting in poor performance in presence of redundancy, and that information about redundancy may be obtained from the PWGC pattern.
We develop here a strategy to combine the two approaches: some
links inferred from PWGC are retained and added to those obtained from CGC. The
PWGC links that are discarded are those which can be
derived as indirect links from the CGC pattern. In the following we
describe the proposed approach in detail.

Let $\Delta$ be the matrix of influences from CGC (or PCGC). Let
$\Delta^*$ be the matrix from PWGC. Non-zero elements of $\Delta$
and  $\Delta^*$ correspond to the estimated influences. Let these
matrices be evaluated using a model of order $m$.

\begin{figure}[ht]
\includegraphics[width=13cm]{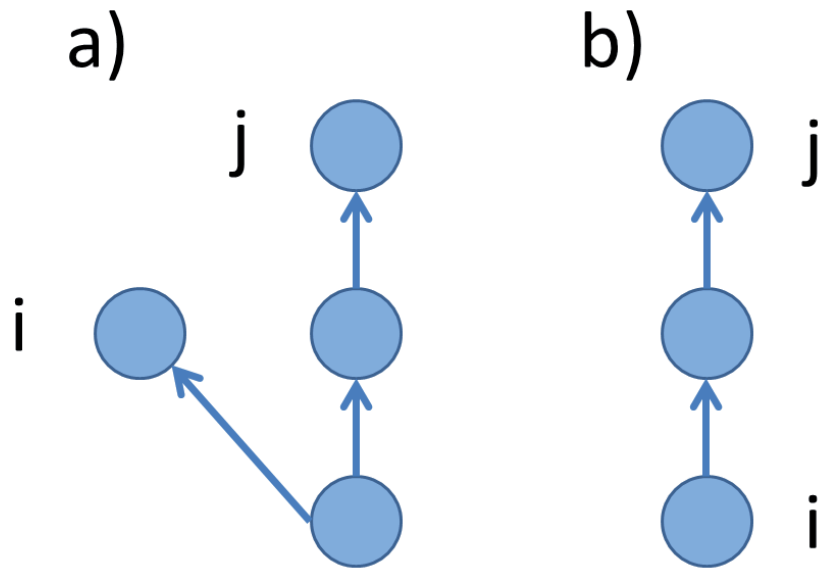}
\caption{{\rm (a) The indirect influence $i\to j$ corresponding to a
nonzero element of the matrix $\Delta^2 \Delta^\top$. If $\Delta^2
\Delta^\top (i,j)\ne 0$, then a common source influences $i$ and $j$
but with different lags. (b) The indirect causality $i\to j$
corresponding to nonzero elements of the matrix $\Delta^2$. If
$\Delta^2 (i,j)\ne 0$, then a third node acting as a mediator of the
interaction $i\to j$.
 \label{fig6}}}\end{figure}

The matrix $$M_{\alpha \beta}=\Delta^\alpha (\Delta^\top)^\beta$$
contains paths of length $\alpha + \beta$ with delays in the range
$[-\beta m +\alpha,\ldots, -\beta + \alpha m]$. Indeed:

\begin{equation}\label{path}M_{\alpha \beta}(i,j)=\sum_{i_1} \sum_{i_2} \cdots
\sum_{i_{\alpha+\beta -1}} \Delta(i,i_1)\Delta(i_1,i_2)\cdots
\Delta(j,i_{\alpha+\beta-1});
\end{equation}
since all elements of $\Delta$ are non-negative, it follows that
$M_{\alpha \beta}(i,j)$ is not vanishing if and only if  it is
possible, in matrix $\Delta$,  to go from node i to
node j moving $\beta$ steps backward and $\alpha$ steps forward,
where a step is allowed if the corresponding element of $\Delta$ is
not zero. Therefore the nonzero elements of the matrix $M_{\alpha \beta}$
describe a situation where two nodes receive a common input from a
third node which is $\alpha$ steps backward in time from one node,
and $\beta$ steps backward in time  in time from the other node. In
other words, if the element $M_{\alpha \beta} (i,j)$ does not
vanish, then there exist an indirect interaction between nodes $i$
and $j$ due to a common input. The circuit corresponding to
$M_{2 1}$ is represented in figure \ref{fig6}a: if the element $M_{2,1} (i,j)$ is non-vanishing, then i and j are connected as in figure \ref{fig6}a .

Since the order of the model is $m$, a simple comparison between the
delays from the common source to $i$ and $j$ demonstrates that the
indirect influence corresponding to the non-zero element
$M_{\alpha,\beta}(i,j)$ might be detected by PWGC only if
$$[-\beta m +\alpha,\ldots, -\beta + \alpha m]\cap [1,\ldots,m] \neq
\emptyset;$$ this is equivalent to $${\beta +1\over m}\le \alpha\le
(\beta +1)m.$$

Now, the matrix $F_\alpha=\Delta^\alpha,$ with  $\alpha \ge 1$,
contains paths of length $\alpha$ with delays in the range
$[\alpha,\ldots, \alpha m]$, indeed:
\begin{equation}\label{path2}F_{\alpha}(i,j)=\sum_{i_1} \sum_{i_2} \cdots
\sum_{i_{\alpha -1}} \Delta(i,i_1)\Delta(i_1,i_2)\cdots
\Delta(i_{\alpha -1},j);
\end{equation}

 Any nonzero element of the matrix
$F_{\alpha}(i,j)$ describes  an indirect causal interaction between nodes
$i$ and $j$ where $i$ sends information to $j$ through a cascade of
$\alpha$ links: $i\to i_1$, $i_1\to i_2$, \ldots, $\i_{\alpha -1}
\to j$. The circuit corresponding to $F_{2}$ is depicted in figure
\ref{fig6}b. The indirect causal interaction $i\to j$, corresponding
to the non-zero element $F_\alpha (i,j)$ might be detected by PWGC
 if $\alpha \le m$.

Let us now consider the matrix $$B=\sum_{\alpha,\beta} M_{\alpha
\beta} +\sum_{\alpha^\prime =1}^m  F_{\alpha^\prime},$$ where the
first sum is over pairs $\{\alpha, \beta\}$ satisfying ${\beta
+1\over m}\le \alpha\le (\beta +1)m.$ If $B_{ij}$ is non vanishing,
then according to CGC there is an indirect causal interaction
between $i$ and $j$: therefore PWGC might misleadingly reveal such interaction
considering it a direct one. In the approach just described we discard (as indirect) the links found by
PWGC for which $B(i,j)\ne 0$. Therefore in the
pairwise matrix $\Delta^*$ we set to zero all the elements such that
$B_{ij}>0$ (pruning). The resulting matrix $\Delta^*$ contains links
which cannot be interpreted as indirect links of the multivariate
pattern, and will be retained and ascribed to redundancy effects.

For $m=1$ we have that the only terms in the first sum are those
with $\alpha = \beta +1$, so the first non trivial terms are
$$B_1=\Delta+\Delta^2 \Delta^\top.$$ For $m=2$, the simplest terms are:
$$B_2=\Delta+ \Delta^2+\Delta \Delta^\top.$$
Since, due to the finite number of samples, a mediated interaction
is more unlikely to be detected (by the pairwise analysis) if it
corresponds to a long path, we limit the sum in the matrix $B$ to
the simplest terms.

\begin{figure}[ht]
\includegraphics[width=13cm]{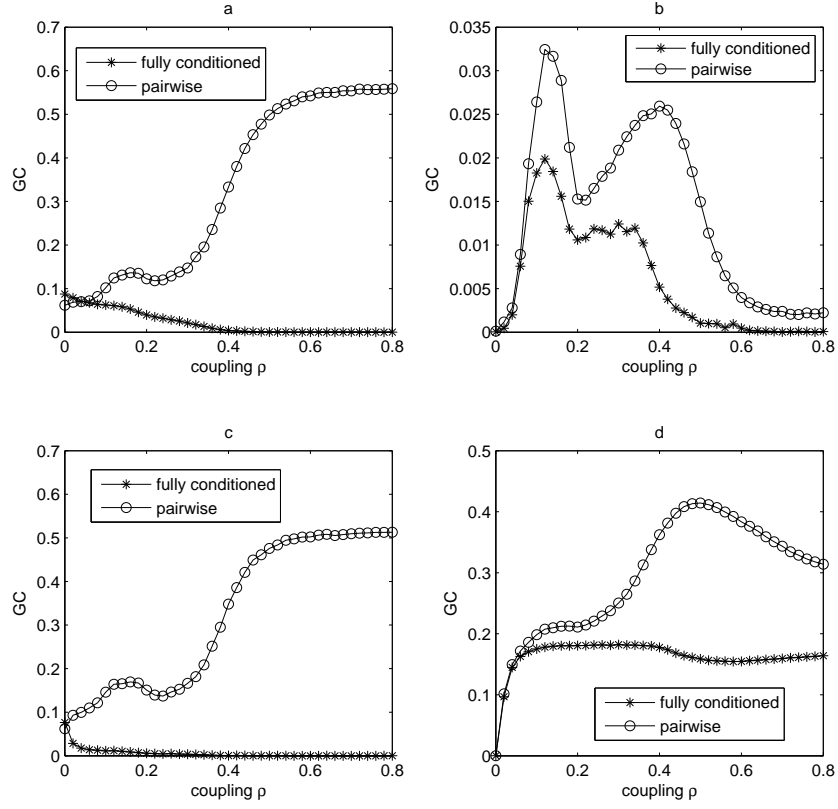} 
\caption{{\rm (a) The CGC
and PWGC of the causal interaction $x_1 \to x_5$ is plotted as a
function of the coupling $\rho$ for the toy model for the proposed
approach combining PWGC and CGC. The linear algorithms are used here
and results are averaged over 100 runs of 2000 time points (b) The
CGC and PWGC of the causal interaction between two variables in the
multiplet is plotted as a function of the coupling $\rho$ for the
toy model for the proposed approach combining PWGC and CGC. The
linear algorithms are used here and results are averaged over 100
runs of 2000 time points (c) The nonlinear CGC and PWGC of the
causal interaction $x_1 \to x_5$ is plotted as a function of the
coupling $\rho$ for the toy model for the proposed approach
combining PWGC and CGC. The algorithm with the polynomial kernel of
order 2 is used here and results are averaged over 100 runs of 2000
time points (d) The nonlinear CGC and PWGC of the causal interaction
between two variables in the multiplet is plotted as a function of
the coupling $\rho$ for the toy model for the proposed approach
combining PWGC and CGC. The algorithm with the polynomial kernel of
order 2 is used here and results are averaged over 100 runs of 2000
time points. \label{fig7}}}\end{figure}

As a toy example to illustrate an application of the proposed approach, we
consider a system made of five variables $\{x_i\}$. The first four
constitute a multiplet made of an unidirectionally coupled logistic
maps,  eqs.(\ref{catena1}-\ref{catena2}) with $i$ ranging in
$\{1,2,3,4\}$, coupling $\rho$ and interactions $1\to 2$, $2\to 3$
and $3\to 4$. The fifth variable is influenced by the mean field
from the coupled map lattice, see equation (\ref{synch2}). The four
variables in the multiplet become {\it almost} synchronized for
$\rho
>0.4$. In figure \ref{fig7} we depict both
the average influence from the variables in the multiplet to $x_5$,
and the average influence between pairs of variables in the
multiplet: both linear and nonlinear PWGC and CGC are shown for the
two quantities. Note that only the nonlinear algorithm  correctly
evidences the causal interactions within the multiplet of four
variables, whilst the linear algorithm detects a very low causal
interdependency among them. The driving influence from
the multiplet to $x_5$ detected by CGC vanishes at high coupling
redundancy, both in the linear and nonlinear approach, due to the redundancy induced by synchronization.

To explain how the proposed approach works we describe two
situations, corresponding to low and high coupling. At low coupling, the CGC approach
estimates the correct causal pattern in the system, and the nonzero
elements of $\Delta$ are $1\to 2$, $2\to 3$, $3\to 4$, $1\to 5$,
$2\to 5$, $3\to 5$ and $4\to 5$. The nonzero elements of the matrix $\Delta^*$, from PWGC
analysis, are the same as $\Delta$ plus $1\to 3$, $1\to 4$, $2\to 4$,
corresponding to indirect causalities; however these three
interactions lead to non-zero elements of $\Delta^2$ (and,
therefore, of $B$), hence they must be discarded. It follows that
$\Delta^*$ does not provide further information than $\Delta$ at low coupling.

On the contrary at high coupling, due to synchronization, the CGC
approach does not reveal the causal interactions $1\to 5$, $2\to 5$,
$3\to 5$ and $4\to 5$, whilst still they are recognized by PWGC;
$\Delta^*$ is still nonzero in correspondence of $1\to 5$, $2\to 5$,
$3\to 5$ and $4\to 5$, while the corresponding elements of $B$ are
vanishing. According to our previous discussion, the interactions
$1\to 5$, $2\to 5$, $3\to 5$ and $4\to 5$, detected by PWGC, should not be
discarded: combining the results by CGC and PWGC we obtain the
correct causal pattern even in presence of strong synchronization.

\section{Application to gene expression data.}
\begin{figure}[ht]
\includegraphics[width=13cm]{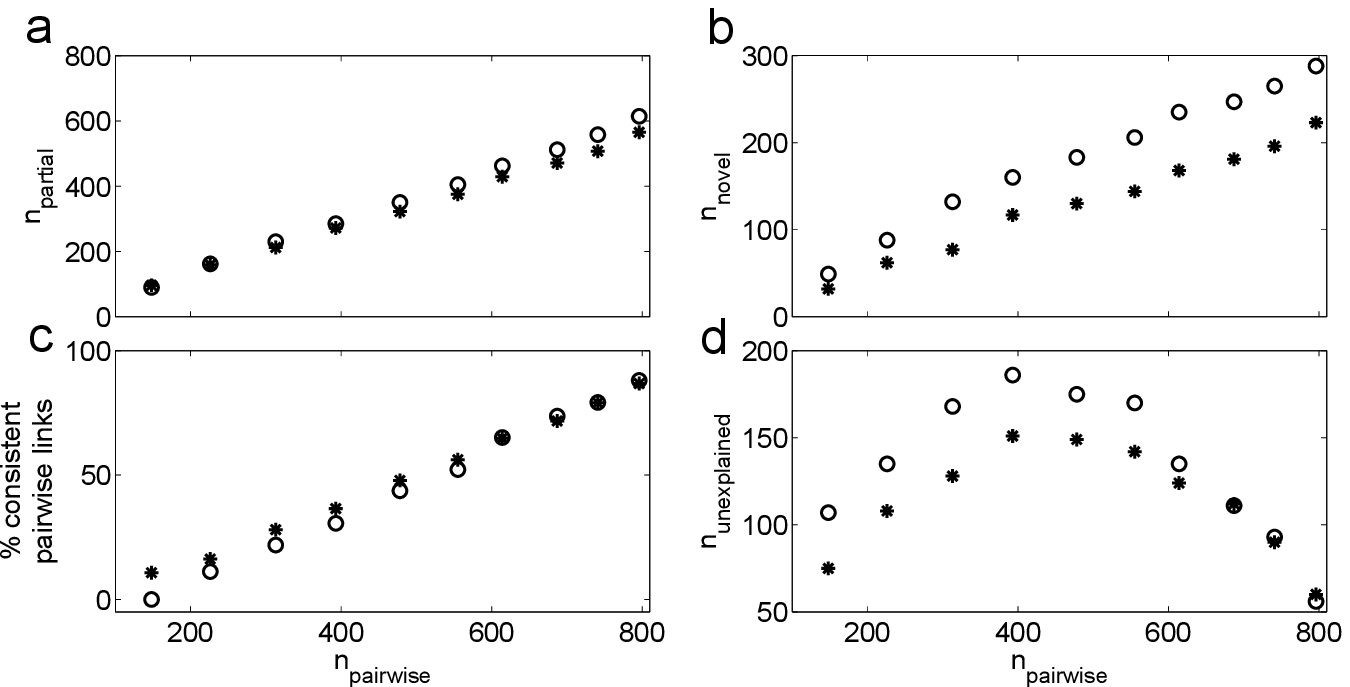}
\caption{{\rm Concerning the genetic application, several quantities
are plotted as a function of the number of bivariate causality links
exceeding the threshold. (a) The number of retrieved links by PCGC
with strategies IB ($\ast$) and PB (o). (b)  The number of retrieved
links by PCGC, with strategies IB ($\ast$) and PB (o), which are not
present in the bivariate pattern. (c) The percentage of retrieved
links, by BVGC, which are consistent with the PCGC with strategies
IB ($\ast$) and PB (o). (d) The number of retrieved links by BVGC,
which are not consistent with PCGC ( with strategies IB ($\ast$) and
PB (o)).
 \label{fig8}}}\end{figure}

HeLa \cite{hela} is a famous cell culture, isolated from a human
uterine cervical carcinoma in 1951. HeLa cells have acquired
cellular immortality, in that the normal mechanisms of programmed
cell death after a certain number of divisions have somehow been
switched off. We consider the HeLa cell gene expression data of
\cite{fujita}. Data corresponds to $94$ genes and $48$ time points,
with an hour interval separating two successive readings (the HeLa
cell cycle lasts 16 hours). The 94 genes were selected from the full
data set described in \cite{whitfield}, on the basis of the
association with cell cycle regulation and tumor development. We
apply linear PWGC and linear CGC (using just another conditioning
variable, and using both the selection strategies IB and PB
described in Section III). We remark that the CGC approach is
unfeasible in this case due to the limited number of samples. Due to
the limited number of samples, in this case we do not use
statistical testing for assessing the significance of the retrieved
links, rather we introduce a threshold for the influence and analyze
the pattern as the threshold is varied. In figure \ref{fig8} results
are reported as a function of the number of links found by PWGC
$n_{pairwise}$ (which increases as the threshold is decreased); we
plot (1) the number of links found by PGC $n_{partial}$, (2) the
number of links found by PGC and not by PWGC $n_{novel}$, which are
thus a signature of synergy, (3) the percentage of pairwise links
which can be explained as direct or indirect causalities of the PGC
pattern (thus being consistent with the partial causality pattern),
found using the matrix $B_1$ to detect the indirect links, which
correspond to circuits like the one described in figure \ref{fig6}a,
(4)  the number of causality links found by PWGC and not consistent
with PWGC $n_{unexplained}$, corresponding to redundancy. The two
curves refers to the two selection strategies for partial
conditioning.

The low number of samples here allowed us just to use one
conditioning variable, and therefore to analyze only circuits of
three variables; a closely related analysis, see \cite{mindy}, has
been proposed to study how a gene modulates the interaction between
two other genes. On the other hand, the true underlying gene
regulatory network being unknown, we cannot assess the performances
of the algorithms in terms of correctly detected links.

 We  note that both $n_{novel}$ and $n_{unexplained}$ assume
relatively large values, hence both redundancy and synergy
characterize this data-set. The selection strategy PB yields
slightly higher values of $n_{novel}$ and $n_{unexplained}$, emerging then as a better discriminator of synergy and redundancy than IB. A comparison iwth the fully conditioned approach is not possible in this case. On the other hand, as far as the search for synergetic
effects is concerned, we find that the synergetic interactions found
by PCGC with the two strategies are not coinciding, indeed only
$10\%$ of all the synergetic interactions are found by both
strategies. This suggests that when searching for suppressors,
several sets of conditioning variables should be used in CGC in
order to explore more possible alternative pathways, especially when
there is not a priori information on the network structure.

\section{Conclusions}
In this paper we have considered the inference, from time series
data, of the information flow between subsystems of a complex
network, an important problem in medicine and biology. In particular
we have analyzed the effects that synergy and redundancy induce on
the Granger causal analysis of time series.

Concerning synergy, we have shown that the search for synergetic
contributions in the information flow is
equivalent to the search for suppressors, i.e. variables that
improve the predictive validity of another variable. Pairwise
analysis fails to put in evidence this kind of variables; fully
multivariate Granger causality solves this problem: conditioning on
suppressors variables leads to nonzero Granger causality. In cases
when the number of samples is low, we have shown that partially
conditioned Granger causality is a valuable option, provided that
the selection strategy, to choose the conditioning variables,
succeeds in picking the suppressors. In this paper we have
considered two different strategies: choosing the most informative
variables for the candidate driver node, or choosing the nodes with
the highest pairwise influence to the target. From the several
examples analyzed here we have shown that the  first strategy is
viable in presence of redundancy, whilst when the interaction
pattern has a tree-like structure, the latter is preferable; however the issue of
selecting variables for partially conditioned Granger causality
deserves further attention as it corresponds to the search for
suppressor variables and correspondingly of synergetic effects. We
have also provided evidence, for the first time, that synergetic
effects are present in an epileptic brain in the preictal condition
(just before the seizure).

We have then shown that fully conditioned Granger approaches do not
work well in presence of redundancy. To handle redundancy, we
propose to split the pairwise links in two subsets: those which
correspond to indirect connections of the multivariate Granger
causality, and those which are not. The links that are not explained
as indirect connections are ascribed to redundancy effects and they
are merged to those from CGC to provide the full causality pattern
in the system. We have applied this approach to a genetic data-set
from the HeLa culture, and found that the underlying gene regulatory
networks are characterized by both redundancy and synergy, hence
these approaches are promising also w.r.t. the reverse engineering
of gene regulatory networks.

In conclusion, we observe that the problem of inferring reliable
estimates of the causal interactions in real dynamical complex
systems, when limited a priori information is available, remains a major theoretical challenge.
In the last years the most important results in this
direction are related to the use of data-driven approaches like Granger
causality and transfer entropy. In this work we have shown that in
presence of redundancy and synergy, combining the results from the
pairwise and conditioned approaches may lead to more effective analyses.

\clearpage

\clearpage

\clearpage

\clearpage

\clearpage

\clearpage

\clearpage

\clearpage


\begin{thebibliography}{99}
\bibitem{ivanov} A. Bashan, R.P. Bartsch, J. W. Kantelhardt, S. Havlin, P. Ch. Ivanov, Nature Communications 3, 702, (2012)
\bibitem{barabasi} A.L. Barabasi, {\em Linked: the new science of networks}. (Perseus Publishing, Cambridge Mass., 2002);
S. Boccaletti, V. Latora, Y.  Moreno, M. Chavez and D.-U.  Hwang,
Phys. Rep. {\bf 424}, 175 (2006).
\bibitem{netnn}L.F. Abbott, C. van Vreeswijk, Phys. Rev. {\bf E 48},
1483 (1993).
\bibitem{gennet}T.S. Gardner, D. Bernardo, D. Lorenz, J.J. Collins,
Science {\bf 301}, 102 (2003).
\bibitem{protein} C.L. Tucker, J.F. Gera, P. Uetz, Trends Cell Biol.
{\bf 11}, 102 (2001).
\bibitem{metabolic}H. Jeong, B. Tombor, R. Albert, Z.N. Oltvai, A.L.
Barabasi, Nature {\bf 407}, 651 (2000).
\bibitem{metabolic1}R. Guimerà, L.A. Nunes Amaral,
Nature {\bf 433}, 895 (2005)
\bibitem{cortes2011}
IM De la Fuente, JM Cortes, MB Perez-Pinilla, V Ruiz-Rodriguez and J Veguillas, PLoS One {\bf 6}, e27224 (2011)
\bibitem{pereda2005}
E. Pereda, R. Quiroga, J. Bhattacharya, Progress in Neurobiology
{\bf 77}, 1 (2005)

\bibitem{book1}M. Wibral, R. Vicente and J.T. Lizier (Eds.) {\em Directed Information Measures in Neuroscience}
 (Springer, Berlin, 2014);
\bibitem{book2}K. Sameshima, L.A. Baccala (Eds.) {\em Methods in Brain Connectivity Inference through Multivariate Time Series Analysis}
 (CRC press, 2014);
\bibitem{granger} C.W.J. Granger, Econometrica {\bf 37}, 424 (1969).
\bibitem{sethbressler}
S.L. Bressler, A.K.  Seth, Neuroimage {\bf 58}, 323 (2011)
\bibitem{schreiber} T. Schreiber, Phys. Rev. Lett.
{\bf 85}, 461 (2000).
\bibitem{barnett}L. Barnett, A. B. Barrett, and A. K. Seth,  Phys.
Rev. Lett., vol. 103, no. 23, 2009.
\bibitem{hac} K. Hlavackova-Schindler, Applied Mathematical Sciences {\bf 5}, 3637 (2011)
\bibitem{vicente}R. Vicente, M. Wibral, M. Lindner, G. Pipa, J Comput Neurosci {\bf 30}, 45 (2011)
\bibitem{njp} D. Zhou, Y. Zhang, Y. Xiao and D. Cai, New J. Phys. {\bf 16}, 043016 (2014)
\bibitem{bart1} R. P. Bartsch, P. Ch. Ivanov,
Communications in Computer and Information Science, 438, pp 270-287 (2014)
\bibitem{bart2} R. P. Bartsch, A. Y. Schumann, J. W. Kantelhardtd, T. Penzele, and P. Ch. Ivanov,
Proceedings of the National Academy of Sciences 109 (26), 10181-10186 (2012)
\bibitem{njp1}T. Ge, Y. Cui, W. Lin, J. Kurths and C. Liu,  New J. Phys. {\bf 14} 083028 (2012)
\bibitem{geweke} J. F. Geweke,
Journal of the American Statistical Association, vol. 79, no. 388,
pp. 907-915, 1984.
\bibitem{partialold} D. Marinazzo, M. Pellicoro, and S. Stramaglia,
Computational and Mathematical Methods in Medicine, Volume 2012
(2012), Article ID 303601.
\bibitem{partialBC} G. Wu, W. Liao, H. Chen, S. Stramaglia, D. Marinazzo
Brain Connectivity, 3(3): 294-301 (2013)
\bibitem{noipre} L. Angelini et al., Phys. Rev. {\bf E 81}, 037201
(2010).
\bibitem{griffith}V. Griffith  and C. Koch, (2014). �Quantifying synergistic mutual information,� in Guided Self-Organization: Inception, Vol. 9, ed. M. Prokopenko (Berlin: Springer), 159�190.
\bibitem{harder}M. Harder, C. Salge and D. Polani, Phys. Rev. E {\bf 87}, 012130 (2013)
\bibitem{Lizier} J.T. Lizier, B. Flecker, P.L. Williams Artificial Life (ALIFE), 2013 IEEE Symposium on , pp.43,51, doi: 10.1109/ALIFE.2013.6602430 (2013)
\bibitem{yang}S. Yang, J. Gu, Jour of Zhejiang University Science {\bf
5}, 1382 (2004)
\bibitem{schnei} E. Schneideman, W. Bialek, M.J. Berry, J Neurosci
{\bf 23}, 11539 (2003)
\bibitem{gat}I. Gat and N. Tishby, NIPS page 111-117, the MIT press
(1998).
\bibitem{barnettpre}A. B. Barrett, L. Barnett, and A. K. Seth,  Phys.Rev. E, vol. 81,
no. 4, Article ID 041907, 14 pages, 2010.
\bibitem{zhou} Z. Zhou, Y.
Chen, M. Ding, P. Wright, Z. Lu, and Y. Liu,  Human Brain Mapping,
vol. 30, no. 7, pp. 2197 (2009)
\bibitem{noipla}D. Marinazzo,
W. Liao, M. Pellicoro, and S. Stramaglia,  Physics Letters, Section
A, vol. 374, no. 39, pp. 4040  (2010)
\bibitem{conger} AJ Conger, Educational and Psychological Measurement April 1974 vol. 34 no. 1 35-46
\bibitem{thompson} FT Thompson, DU Levine, Multiple Linear Regression
Viewpoints, 24: 11 - 13 (1997)
\bibitem{preExpansion} S. Stramaglia, G. Wu, M. Pellicoro, D.
Marinazzo Physical Review E, 86,
066211 (2012)
\bibitem{Chicharro} D. Chicharro, A. Ledberg Physical Review E, 86,
041901 (2012)
\bibitem{hla} K. Hlavackova-Schindler, M. Palus, M. Vejmelka, J.
Bhattacharya, Physics Reports {\bf 441}, 1 (2007).
\bibitem{noipre2}D. Marinazzo, M. Pellicoro and S. Stramaglia, Phys. Rev. E {\bf 77}, 056215 (2008).
\bibitem{vapnik}V. Vapnik. The Nature of Statistical Learning Theory. Springer,
N.Y., 1995.
\bibitem{shawe} J. Shawe-Taylor and N. Cristianini,
{\em Kernel Methods For Pattern Analysis}. (Cambridge University
Press, London, 2004)
\bibitem{noiprl} D. Marinazzo, M. Pellicoro, S. Stramaglia,
Phys. Rev. Lett. {\bf 100}, 144103 (2008).
\bibitem{ancona} N. Ancona and S. Stramaglia, Neural Comput. {\bf 18}, 749 (2006).
\bibitem{kol_web} http://math.bu.edu/people/kolaczyk/datasets.html, accessed may
2012
\bibitem{kol_paper} M.A. Kramer, E.D. Kolaczyk, H.E. Kirsch,
Epilepsy Research {\bf 79}, 173, 2008
\bibitem{hela}J.R. Masters,
Nature Reviews Cancer {\bf 2}, 315 (2002).
\bibitem{fujita}A. Fujita et al., BMC System Biology {\bf 1:39}, 1
(2007).
\bibitem{whitfield}M.L. Whitfield et al., Mol. Biol. Cell {\bf 13}, 1977
(2002).

%
\bibitem{mindy} K. Wang, et al. Nature Biotechnology 27, 829–837 (2009) doi:10.1038/nbt.1563

\end{thebibliography}
\end{document}